# Multi-field nanoindentation apparatus for measuring local mechanical properties of materials in external magnetic and electric fields

Hao Zhou,[1] Yongmao Pei,[1,a)] Hu Huang,[2] Hongwei Zhao,[2] Faxin Li,[1] and Daining Fang[1,a)]

[1] *State Key Laboratory for Turbulence and Complex Systems, College of Engineering, Peking University, Beijing 100871, China*
[2] *College of Mechanical Science & Engineering, Jilin University, Changchun, Jilin 130025, China*

Nano/micro-scale mechanical properties of multiferroic materials can be controlled by the external magnetic or electric field due to the coupling interaction. For the first time, a modularized multi-field nanoindentation apparatus for carrying out testing on materials in external magnetostatic/electrostatic field is constructed. Technical issues, such as the application of magnetic/electric field and the processes to diminish the interference between external fields and the other parts of the apparatus, are addressed. Tests on calibration specimen indicate the feasibility of the apparatus. The load-displacement curves of ferromagnetic, ferroelectric and magnetoelectric materials in the presence/absence of external fields reveal the small-scale magnetomechanical and electromechanical coupling, showing as the $\Delta E$ and $\Delta H_{in}$ effects, i.e. the magnetic/electric field induced changes in the apparent elastic modulus and indentation hardness.

## I. INTRODUCTION

Nanoindentation technique has been extensively used in determining the elastic modulus and hardness of materials,[1-2] and investigating the nanoscale contact phenomena, such as the anelasticity of ferroelectrics,[3] the earlier than expected onset of plasticity of metals,[4] and the phase transition of semiconductors[5]. However, magnetostrictive and piezoelectric materials are usually employed in external mechanical stress and magnetic or electric field as actuators and transducers. On the one hand, the mechanical properties in the presence of external field can be quite different than that without these fields due to the intrinsic magnetomechanical[6] or electromechanical coupling[7]. It is of vital significance to have a better understanding of the external field dependent mechanical deformation behaviour of such materials under nearly the real service conditions. On the other hand, the conventional nanoindentation cannot provide the multi field and meet the request to study the magnetic/electric field controlled mechanical properties.

The experimental setup presented in Ref. 8 can measure the indentation induced current transients of piezoelectrics with a conductive indentor of 500 μm nominal radius, but cannot supply an external electric or magnetic field. The NanoECR instrument co-developed by Hysitron Inc. and Ruffell et al. from Australian National University can be used to measure the contact resistance at different penetration depth according to the voltage-current curves.[9,10] However, there is also no magnetism module. The experimental facility described in Ref. 11 is based on the standard microhardness tester, which can measure the microhardness at a constant load under the superposition of constant magnetic field and an impulse electrical current, but it cannot obtain the loading/unloading curves, the contact stiffness, and the elastic modulus. What's more, the thermal and electromagnetic interference resulted from the electromagnet makes it almost impossible to conduct a high-precision nanoindentation testing in an adjusting external electromagnetic field. Up to now, there is no real multi-field nanoindentation apparatus reported.

In order to probe the field dependent mechanical and electromechanical/magnetomechanical properties in the small scale, a multi-field nanoindentation apparatus has been designed and constructed, which allows the simultaneous application of penetration load/displacement, DC voltage, and magnetic field. Therefore, the multi-field nanoindenter developed by us enables the measurement of load-depth curves, indentation hardness, contact stiffness and reduced modulus at both nano-scale and micro-scale in external magnetic or electric field. Experimental results, including the magnetic field dependent load-depth curves for ferromagnetic Ni single crystal, the electric field dependent curves for ferroelectric PMN-PT single crystal, as well as the magnetic and electric fields dependent curves for layered magnetoelectric composite $La_{0.7}Sr_{0.3}MnO_3$/PIN-PMN-PT are presented to show the main functions and usability of the measuring system. At the same time, they reveal the small-scale magnetomechanical and electromechanical coupling effect, showing as the $\Delta E$ and $\Delta H_{in}$ effects.

## II. MODULAR DESIGN AND MULTI-FIELD PERFORMANCE TEST

The apparatus mainly consists of three modules: the mechanical, the magnetic and the electrical modules. Specifically, it includes: a specially created magnetic field generation device made up of permanent magnet and soft magnetic materials, a Gauss meter with a Hall effect magnetometer for magnetic field measurement, a DC regulated power supply and a conductive indenter tip to provide a direct electric field across the specimen, a piezoelectric actuator to drive the conductive indenter tip, as well as the precise load and displacement sensors to measure the indentation force and depth.

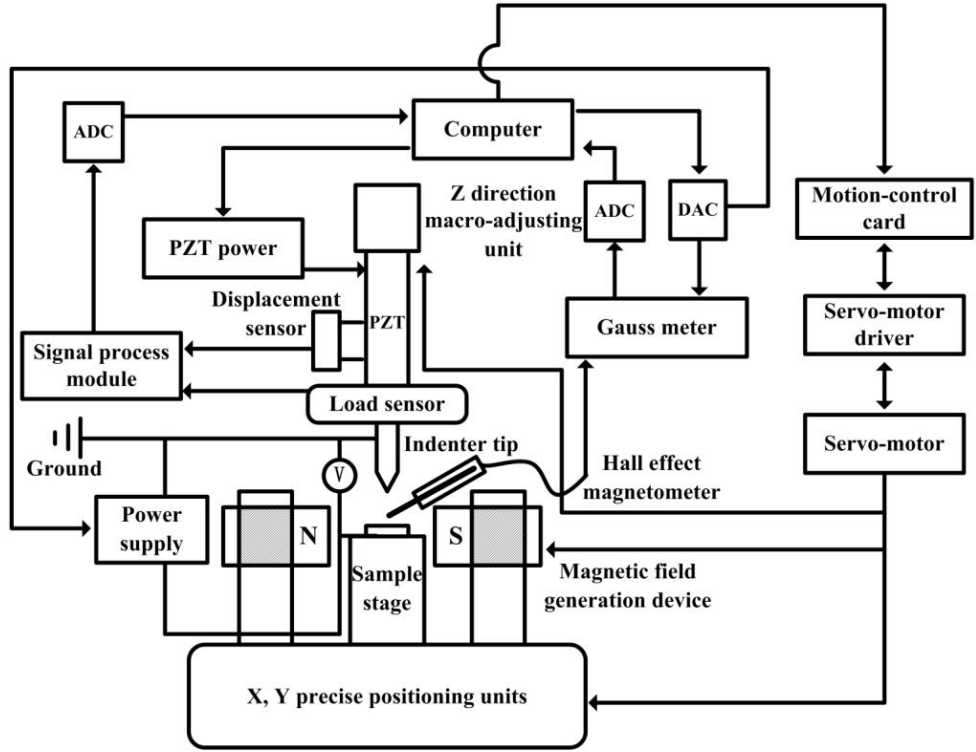

FIG. 1. Schematic diagram of the multi-field nanoindenter.

The schematic diagram of the magneto-electro-mechanical loading and measuring system with control and measurement components is shown in Fig. 1. The variable magnetic field is applied to the sample by changing the distance of the magnetic poles. The magnetic field strength is measured with the Gauss meter and Hall sensor, and the signal is transmitted to the computer through the analog/digital converter. The electric field is supplied to the sample by a voltage source controlled by the computer via a digital/analog converter. The indentation load is provided by the piezoelectric actuator (PZT) through closed-loop control, and the displacement and force are acquired by the linear variable differential transformer (LVDT) and the strain load sensor. The collected experimental data are processed with the computer, and then the parameters of the sample can be obtained.

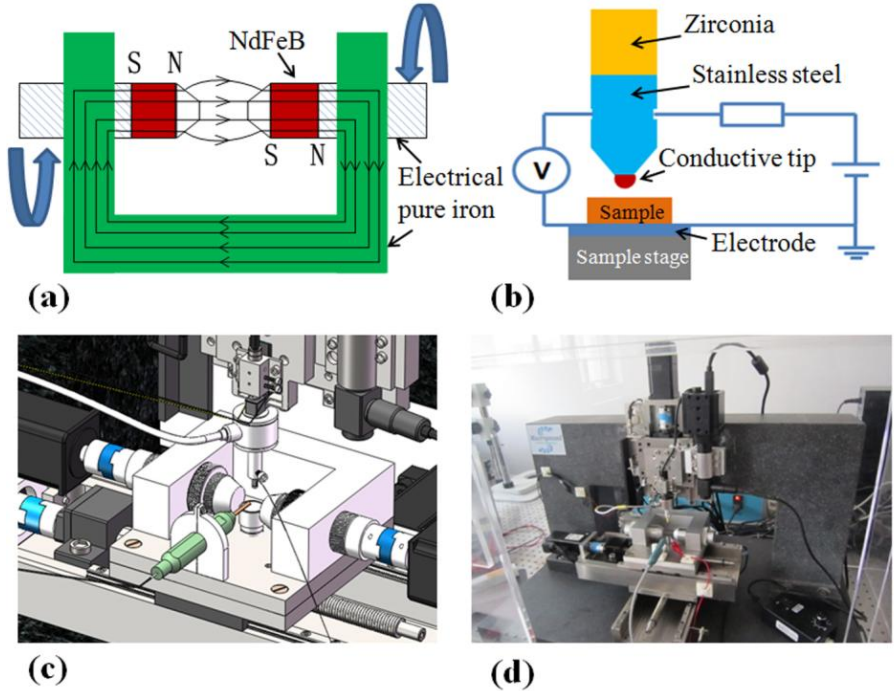

FIG. 2. Drawings and photo of the actual apparatus: (a) schematic of the magnetic field generation device; (b) schematic of the electric field module; (c) three-dimensional renderings for the central part of the multi-field nanoindenter; (d) the photo of the actual apparatus.

The magnetic field generation device is illustrated in Fig. 2a. It consists of two cylindrical rare-earth permanent magnets NdFeB (the red part with N and S poles) with a cylindrical bar (the dashed part) and a cone frustum pole (the white part) made of electrical pure iron glued to their two ends. A U-shape magnet yoke made of iron was connected to the iron cylindrical bar with threaded connection. One reason is to reduce the magnetic resistance of the circuit so as to enhance the magnetic field strength around the sample, and the other reason is to diminish the leakage flux in the space, avoiding its influence on the sensors of the apparatus. The distance between the iron poles can be adjusted from 0~70 mm by rotating the cylindrical bars (the dashed part) driven by the motor under the control of the computer or by manually. The sample is placed in the center of the space between the two poles, where the magnetic field is relatively uniform. The distance between the two iron poles versus the magnetic field strength curve is shown as the line with solid squares in Fig.3. The magnetic field strength can be as large as 8000 Oe when the distance between the poles is less than 3 mm. Even at the distance of 16 mm, the magnetic field strength is over 2000 Oe, which is sufficient to switch the micro structure of most soft ferromagnetic materials. And the magnetic field strength variation is less than 5% in the central $5\times5\times2$ mm$^3$ (length×width×height) space, in which the sample can be placed. As the changing of the magnetic field strength at the sample involves adjusting the distance between the iron poles, the application and measuring resolution of the magnetic field strength at the indentation position is iron poles distance dependent, due to the space inhomogeneity of the magnetic field and the error of locating the indenter tip and the Hall probe. For the magnetic field strength $H$<500 Oe, <700 Oe, <2000 Oe, and >2000 Oe, the resolutions are 1 Oe, 5Oe, 10Oe, and 1%×$H$, respectively. Once a demanding magnetic field is achieved, the motors driving the motion of the iron poles are turned off. Thus there is no distance changing between the two poles resulted from the vibration of the motors during the indentation process, and the magnetic field is stable over time. In order to avoid the influence of magnetic field on the load and displacement sensors, the sensors are installed away from the magnetic field device. Using a permanent magnet design rather than electromagnetic coils, we avoid the unwanted thermal drift resulted from the large current to maintain the magnetic field. The magnetic field strength around the sensors is nearly zero, as the hollow triangles illustrated in Fig. 3.

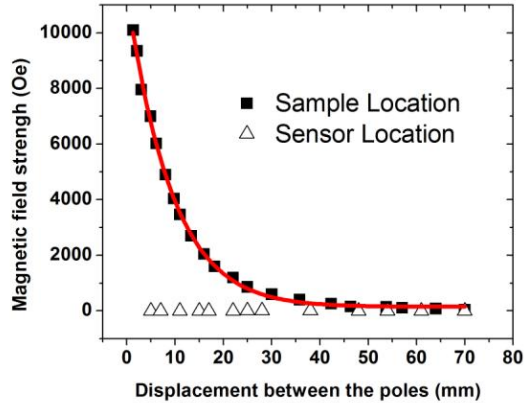

FIG. 3. The magnetic field strength at the center of sample position (solid squares) and around the sensors (hollow triangles) versus the distance between the two iron poles.

Meanwhile, in order to avoid the electric field's effects on the sensors, the indenter shaft is composed of the upper insulating segment (zirconia) and the lower conducting segment (stainless steel) with a groove that can be connected to the voltage source (Maximum voltage: 1000 V; Resolution: 1.5 V) via a wire. On the tip of the shaft, the Berkovich indenter probe (from SYNTON-MDP) manufactured from boron doped diamond is conductive with a radius of curvature of 150 nm, as can be seen in Fig. 2b. Besides, a CCD camera (400× magnification) is used to help the location of the indentation before testing and observe the impression morphology after testing. The apparatus is placed on the vibration isolation platform to reduce the environmental vibration interference.

In order to avoid possible electromagnetic interference, the piezoelectric actuator is used to drive the indentation process instead of the traditional electromagnetic force loading approach adopted by certain instrumented indenter manufacturers. The performance parameters of the mechanical module of the apparatus are provided in brief: Maximum load: 1N; Load noise: 35μN; Maximum depth: 20μm; Depth noise: 8nm.

To test the performance of the modularized multi-field nanoindentation apparatus, two pieces of fused quartz from Hysitron Inc. are used as the calibration samples. During the magneto-nanoindentation calibration, the fused quartz sample is glued to a nonmagnetic sample stage to avoid the distortion of the magnetic field. Nanoindentation tests are conducted at various magnetic field strengths. The reduced modulus $E_r$ and indentation hardness $H_{in}$ can be obtained according to the Oliver-Pharr method.[1]

$$E_r = \frac{\sqrt{\pi}}{2} \frac{S}{\sqrt{A}}, \tag{1}$$

$$H_{in} = \frac{P_{max}}{A}, \tag{2}$$

where $S = dP/dh$ is the experimentally measured stiffness of the upper portion of the unloading curve, $A$ is the projected area of the elastic contact, and $P_{max}$ is the maximum indentation load. The reduced modulus can be used to determine the elastic modulus of the sample via

$$\frac{1}{E_r} = \frac{1-\nu^2}{E} + \frac{1-\nu_i^2}{E_i}. \tag{3}$$

Here, $E$ and $\nu$ are modulus and Poisson ratio of the sample, while $E_i$ and $\nu_i$ are modulus and Poisson ratio of the indenter tip.

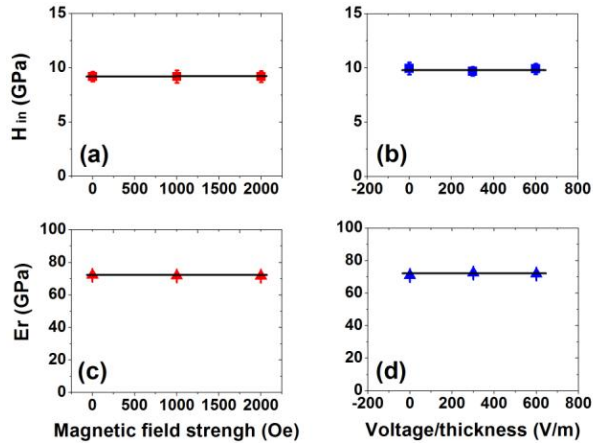

FIG. 4. Calibration for the apparatus in external field with the fused quartz sample: (a,c) the indentation hardness and reduced modulus in external magnetic field; (b,d) the indentation hardness and reduced modulus in external electric field.

It can be seen from Fig. 4(a,c) that the hardness and the reduced modulus are independent of external magnetic field strengths, which indicates the feasibility of the magneto-nanoindentation function. Each data in the figure represents the average of five test values with the error bar showing the standard deviation. These values in Fig. 4 identify well with the field-free data provided by the sample supplier: $9.25 GPa \pm 10\%$ and $69.6 GPa \pm 5\%$ for the hardness and reduced modulus, respectively.

During the electro-nanoindentation calibration, the fused quartz sample is glued to a copper electrode with the conducting resin. The copper electrode is electrically connected to the positive pole of the power source via a wire, while the conducting nanoindentation probe is connected to the negative pole of the power source via another wire. As we can see from Fig. 4(b,d), the electric field also has no effect on the nanoindentation results, indicating the validity of the electro-nanoindentation performance. The same with the above, each data in Fig. 4(b,d) represents the statistics of five tests.

### III. TYPICAL RESULTS

Based on the calibrated multi-field nanoindenter, three kinds of materials (i.e. ferromagnetic, ferroelectric and magnetoelectric materials) are chosen to test the basic function of the developed apparatus, in consideration of their intrinsic magneto-mechanical, electro-mechanical and magneto-electro-mechanical coupling properties.

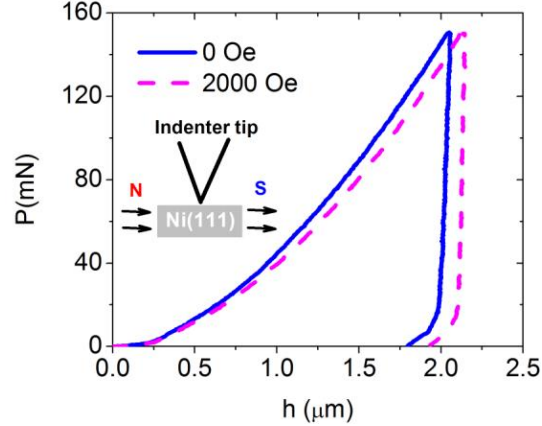

FIG. 5. The indentation load-depth curves of Ni(111) at different magnetic field strength.

A single crystal Ni(111) sample is attached to a non-magnetic alloy stage with the cyanoacrylate adhesive. The non-magnetic property of the adhesive and its high adhesion strength with the sample exclude its contribution to the distortion of a magnetic field and the peeling of the sample from the stage by the magnetic force. The indentations are performed with magnetic field strengths of zero and 2000 Oe, which correspond to the demagnetization state and magnetic saturation state of nickel. Fig.5 shows that the maximum depth, the final residual depth, and the slope of the unloading curve have all changed, which indicate the variation of the contact stiffness and the hardness. From 80 tests at different positions of the sample, the average reduced modulus $E_r$ in the absence/presence of external magnetic field are 156.2±25.8 GPa and 214.8±33.5 GPa, respectively. The indentation hardness $H_{in}$ are 1.66±0.16 GPa and 1.54±0.14 GPa, respectively. The average value of the field-free modulus $E$=164.6 GPa calculated with Eq. (3) approximates the literature values 164 GPa and 175 GPa determined also by indentation techniques on nickel samples.[12,13] According to the data obtained above, the field induced changes in $E_r$ and $H_{in}$ are 38% and -7%, respectively. The great changes at small scale may be related with the contact stress induced magnetic domain switching and the external magnetic field induced domain evolution. Therefore, this apparatus can be used to investigate the small scale $\Delta E$ effect[14,15] and the $\Delta H_{in}$ effect, i.e. the magnetic field induced changes in the apparent elastic modulus and indentation hardness. In addition, it's necessary to note that the magnetic field in present apparatus can only be horizontal, i.e. perpendicular to the indentation axis, because the components of the magnetic field in the indentation axis direction can disturb the signals of the load and displacement sensors just above the sample (see Fig. 2c).

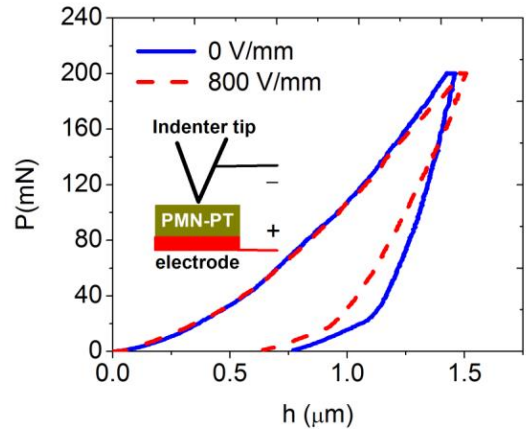

FIG. 6. The indentation load-depth curves of PMN-PT at different electric field strength.

Using an in-plane poled relaxor ferroelectric lead magnesium niobate-lead titanate (PMN-PT) single crystal sample provided by Shanghai Institute of Ceramics, we conduct the nanoindentation tests under DC voltage of zero and 160 V. The thickness of the sample is 0.2 mm. As can be seen from the inset of Fig.6, the positive voltage is applied to the sample through the copper electrode glued to the bottom of the sample with conducting silver paste, while the conductive probe is connected to the negative voltage of the power source. The electric field inside the ferroelectric PMN-PT is inhomogeneous due to the asymmetric top and bottom electrodes. For the high permittivity of the PMN-PT, the voltage drops almost all in the body of PMN-PT, rather than the paste or its interface with the sample. From the curves in Fig. 6, we can see that the unloading slope decreases greatly after the application of electric field, which reflects the obvious decrease of the contact stiffness due to the ferroelectric domains' evolution or the phase boundary moving. What's more, the residual depth also decreases. According to 20 tests, the reduced modulus $E_r$ without/with the application of electric field are 102.6±6.5 GPa and 68.4±4.8 GPa, respectively. The indentation hardness $H_{in}$ are 5.0±0.3 GPa and 6.1±0.3 GPa, respectively. The relative changes in $E_r$ and $H_{in}$ are -33% and 22%, respectively. Elsewhere, the remarkable electric field dependent mechanical properties (manifested in the stress-strain curves) have also been reported by E. A. McLaughlin et al. using the macroscale compression experiment.[7] In present work, to understand the small scale electromechanical response necessitates the knowledge of the electric field distribution inside the sample, however, the inhomogeneous electric field distribution varies with the indenter tip/sample configurations during the indentation process. First, the indenter tip approaches the immediate vicinity of the surface of the sample. Second, the indenter tip is at the contact with the sample surface. Third, the indenter penetrates into the sample to a certain depth. For the first two cases, the electric field distributions can be evaluated utilizing the image charge method[16] or the effective point charge approach[17]. A detailed analysis considering the full dielectric anisotropy has been given by E. A. Eliseev et al.[18] For the third case, the indentation theory of piezoelectric materials and the finite-element method can be used to determine the electric field distributions at different indentation depth.[19] Besides, we can also change the positions of the electrodes, with the indenter tip acting as one or not, to realize the changing of the orientations of the electric field relative to the indentation axis.

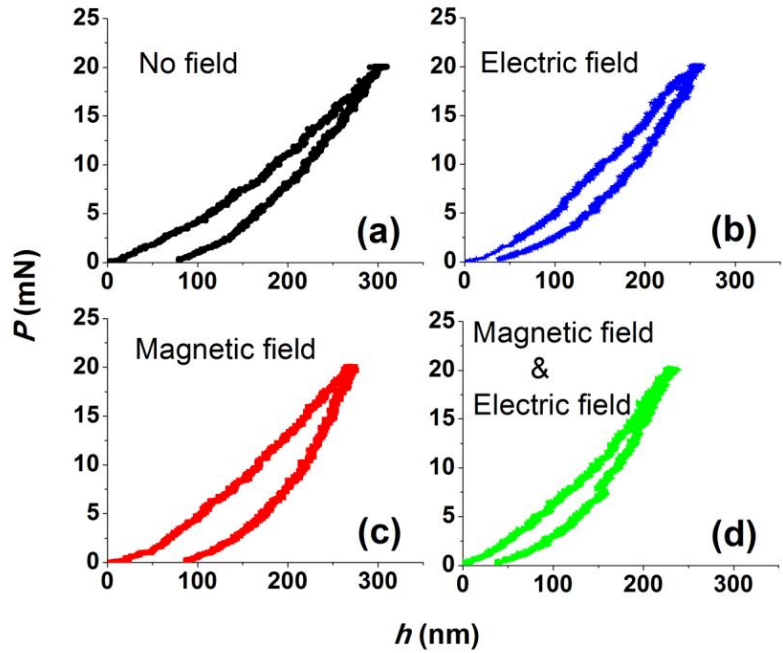

FIG. 7. The indentation load-depth curves of LSMO/PIN-PMN-PT at different electric and magnetic field strength. (a) without external field; (b) with electric field 750 V/mm; (c) with magnetic field 2000 Oe; (d) with both electric field 750 V/mm and magnetic field 2000 Oe.

In recent years, the versatile magnetoelectric materials have attracted more and more attention for their potential applications in multifunctional devices.[20,21] Here, a layered magnetoelectric composite $La_{0.7}Sr_{0.3}MnO_3$/0.33PIN-0.35PMN-0.32PT provided by Shanghai Institute of Ceramics is chosen as an example to illustrate the function of the apparatus, as shown in Fig.7. The magnetic LSMO layer[22] is about 350 nm thick, and the ferroelectric PIN-PMN-PT substrate is 0.4 mm thick. The sample is glued to the copper electrode with conducting silver paste on the substrate side. The copper electrode is connected to the positive pole of the power source while the conducting indenter tip is connected to the negative pole and grounded. The indentation is in the colossal magnetoresistive LSMO layer. Due to the electrical conductivity of LSMO, it also acts as the top electrode of the PIN-PMN-PT layer, and the electric voltage drop is almost all in the ferroelectric PIN-PMN-PT layer. For the geometrical parallel configuration of the top and bottom plate electrodes, the electric field is almost uniform with its orientation in the thickness direction, which attempts to rotate the ferroelectric polarizations to the same direction. The electric poling and the piezoelectric effect can then change the interface misfit strain between the magnetic layer and the ferroelectric layer. What's more, the LSMO itself is a strongly correlated electron system with electro-magneto-structural coupling. All the above factors contribute to the change of the indentation curves' shape with or without external electric/magnetic field. According to five to six tests for each condition (a, b, c, d) in Fig.7, the contact stiffness $S$=137.7$\pm$10.2 mN/µm, 152.8$\pm$4.5 mN/µm, 203.5$\pm$5.9 mN/µm and 179.8$\pm$10.0 mN/µm, respectively. It can be seen that both electric field and magnetic field can increase the stiffness of the composite significantly. Another phenomenon reflected in Fig. 7 is that the electric field makes the residual depth approaching zero, while the magnetic field has little impact on it. What's more, the ratio of irreversible energy loss and total loading work of the whole indentation process also varies, i.e. 0.25$\pm$0.01, 0.11$\pm$0.03, 0.43$\pm$0.01 and 0.24$\pm$0.01 for the four conditions above mentioned, respectively, revealing that the electric field decreases the energy loss while the magnetic field increases it.

It must be pointed out that the load and depth noise floors of the present apparatus are somewhat high in comparison to commercial nanoindenters. This indeed limits the testing of materials in the form of thin films. However, different from the bulk materials, measurements on thin films are more qualitative trend analysis and less accurate quantitative determination of the modulus and hardness. Thus to improve the precision of the apparatus, by improving the signal conditioning module and so on, so as to precisely obtain the mechanical and the multi-field coupling properties of the thin and ultrathin films is the direction of further research.

## IV. CONCLUSIONS

This paper describes a modularized multi-field nanoindenter to study the small scale magneto-mechanical, electro-mechanical, and magneto-electro-mechanical coupling properties. The calibration tests indicate the feasibility of the apparatus. The experiments on ferromagnetic, ferroelectric, and magnetoelectric materials show that the external magnetic or electric field can significantly change the small scale deformation behavior, resulting in the remarkable local $\Delta E$ effect and $\Delta H_{in}$ effect, i.e. the magnetic/electric field induced changes in the apparent elastic modulus and indentation hardness. The newly developed apparatus may be of great significance for the design of the functional devices, in which materials usually work in external magnetic/electric field.

## ACKNOWLEDGMENTS

The authors are grateful for the support by the National Natural Science Foundation of China (11090330, 11090331 and 11072003) and the Chinese National Programs for Scientific Instruments Research and Development (2012YQ03007502). Support by the National Basic Research Program of China (G2010CB832701) is also acknowledged.